\documentclass[manuscript]{aastex}

\usepackage{amsmath}
\usepackage{amssymb}
\usepackage{graphicx}
\usepackage{color}
\usepackage{epstopdf}

\newcommand{\lSect}[1]{{\label{sec:#1}}}

\def\gtaprx {\lower .1ex\hbox{\rlap{\raise .6ex\hbox{\hskip .3ex
    {\ifmmode{\scriptscriptstyle >}\else
        {$\scriptscriptstyle >$}\fi}}}
    \kern -.4ex{\ifmmode{\scriptscriptstyle \sim}\else
        {$\scriptscriptstyle\sim$}\fi}}}
\def\ltaprx {\lower .1ex\hbox{\rlap{\raise .6ex\hbox{\hskip .3ex
    {\ifmmode{\scriptscriptstyle <}\else
        {$\scriptscriptstyle <$}\fi}}}
    \kern -.4ex{\ifmmode{\scriptscriptstyle \sim}\else
        {$\scriptscriptstyle\sim$}\fi}}}
\newcommand{\note}[1]{\emph{\textcolor{black}{}}}

\newcommand{\Msun}{{\ensuremath{M_{\odot}}}}
\newcommand{\Ms}{{\ensuremath{M_{\odot}}}}
\newcommand{\HII}{{\ion{H}{2}}}

\newcommand{\Ni}{{\ensuremath{^{56}\mathrm{Ni}}}}
\newcommand{\Fe}{{\ensuremath{^{56}\mathrm{Fe}}}}

\newcommand{\Co}{{\ensuremath{^{56}\mathrm{Co}}}}
\newcommand{\He}{{\ensuremath{^{4} \mathrm{He}}}}
\newcommand{\Hy}{{\ensuremath{^{1} \mathrm{H}}} }
\newcommand{\Ox}{{\ensuremath{^{16}\mathrm{O}}}}

\newcommand{\Si}{{\ensuremath{^{28}\mathrm{Si}}}}
\newcommand{\Mg}{{\ensuremath{^{24}\mathrm{Mg}}}}
\newcommand{\Cx}{{\ensuremath{^{12}\mathrm{C}}}}

\newcommand{\Sx}{{\ensuremath{^{32}\mathrm{S}}}}

\newcommand{\Ne}{{\ensuremath{^{20}\mathrm{Ne}}}}

\newcommand{\cm}{{\ensuremath{\mathrm{cm}}}}

\newcommand{\erg}{{\ensuremath{\mathrm{erg}}}}

\newcommand{\CASTRO}{\texttt{CASTRO}}
\newcommand{\KEPLER}{\texttt{KEPLER}}

\newcommand{\gcc}{\ensuremath{\mathrm{g}\,\mathrm{cm}^{-3}}}

\makeatletter

\newcommand{\Rmnum}[1]{\expandafter\@slowromancap\romannumeral #1@}
\makeatother

\shorttitle{Collapsed Cores in Globular Clusters}
\shortauthors{Djorgovski et al.}

\begin{document}
\title{General Relativistic Instability Supernova of a Supermassive Population III Star }

\author{ Ke-Jung Chen\altaffilmark{1,2,*}, Alexander Heger\altaffilmark{3}, Stan  
Woosley\altaffilmark{1}, Ann Almgren\altaffilmark{4}, Daniel J. Whalen\altaffilmark{5,6}, and Jarrett L. Johnson\altaffilmark{7} } 

\altaffiltext{*}{IAU Gruber Fellow; kchen@ucolick.org} 

\altaffiltext{1}{Department of Astronomy \& Astrophysics, University of California, Santa 
Cruz, CA 95064, USA} 

\altaffiltext{2}{School of Physics and Astronomy, University of Minnesota, Minneapolis, 
MN 55455, USA}

\altaffiltext{3}{Monash Centre for Astrophysics, Monash University, Victoria 3800, 
Australia} 

\altaffiltext{4}{Center for Computational Sciences and Engineering, Lawrence Berkeley 
National Lab, Berkeley, CA 94720, USA}

\altaffiltext{5}{T-2, Los Alamos National Laboratory, Los Alamos, NM 87545, USA} 

\altaffiltext{6}{Universit\"{a}t Heidelberg, Zentrum f\"{u}r Astronomie, Institut f\"{u}r 
Theoretische Astrophysik, Albert-Ueberle-Str. 2, 69120 Heidelberg, Germany}

\altaffiltext{7}{XTD-PRI, Los Alamos National Laboratory, Los Alamos, NM 87545, USA} 

\begin{abstract}

The formation of supermassive Population III stars with masses $\gtrsim$ 10,000 \Ms\ 
in primeval galaxies in strong UV backgrounds at $z \sim$ 15 may be the most viable 
pathway to the formation of supermassive black holes by $z \sim$ 7.  Most of these 
stars are expected to live for short times and then directly collapse to black holes, with 
little or no mass loss over their lives.  But we have now discovered that non-rotating 
primordial stars with masses close to 55,000 \Ms\ can instead die as highly energetic 
thermonuclear supernovae powered by explosive helium burning, releasing up to 10$
^{55}$ erg, or about 10,000 times the energy of a Type Ia supernova. The explosion is 
triggered by the general relativistic contribution of thermal photons to gravity in the core 
of the star, which causes the core to contract and explosively burn.  The energy release 
completely unbinds the star, leaving no compact remnant, and about half of the mass of 
the star is ejected into the early cosmos in the form of heavy elements.  The explosion 
would be visible in the near infrared at $z \lesssim$ 20 to {\it Euclid} and the Wide-Field 
Infrared Survey Telescope (WFIRST), perhaps signaling the birth of supermassive black 
hole seeds and the first quasars.

\end{abstract}


\section{Introduction}
\lSect{introduction}

The existence of supermassive black holes (SMBHs) in most massive galaxies today 
poses significant challenges to the paradigm of hierarchical structure formation \citep{kormendy1995,ferra2000,ferra2005,gebh2000,matte2005,beif2012,mcco2013}.  In 
particular, it is not known how some SMBHs reached masses of 10$^9$ \Ms\ by $z 
\sim$ 7, less than a Gyr after the big bang \citep{fan2002,fan2006,mort11}.  The two 
leading contenders for the origins of SMBHs are 100 - 500 \Ms\ Population III (Pop III) 
stars at $z \sim$ 20 and supermassive (10$^4$ - 10$^5$ \Ms) Pop III stars at $z \sim$ 
10 - 15 \citep[e.g.,][]{rees1984,madau2001,vol2010, vol12,karlsson2013}.  

The main argument against conventional Pop III stars as SMBH seeds is that their 
BHs must accrete at the Eddington limit continuously down to $z \sim$ 7 to become 
supermassive.  How they could sustain such growth is unclear because they form in 
low-density relic \HII\ regions that delay initial accretion \citep{whalen2004,johnson2007}.  
Later, when they do accrete, the relatively shallow dark matter potential wells in which 
they reside cannot retain their fuel supply because of radiative feedback \citep{awa09}, 
so they grow at well below the Eddington limit \citep{pm11,pm12,pm13}.  Furthermore, 
many low-mass Pop III black holes are ejected from their halos (and thus their fuel 
supply) at birth by asymmetries in their explosion engines \citep{wf12}.

For these reasons, there has been growing interest in supermassive stars (SMSs) as 
candidates for SMBH seeds \citep[e.g.,][]{loeb1994,bromm2003,bege2006,bromm2011,
jet13}.  Such stars could form in atomically-cooled halos at $z \sim$ 15. In this scenario, 
a primeval galaxy forms in a strong Lyman-Werner (LW) UV background that sterilizes 
its constituent halos of H$_2$ prior to assembly \citep{agarw12,jdk12}, preventing them 
from forming stars. When the protogalaxy reaches a mass of $\sim$ 10$^8$ \Ms\ and 
virial temperature of $\sim$ 10$^4$ K, atomic cooling triggers the rapid collapse of gas 
at its center at rates of up to 1 \Ms\ yr$^{-1}$ \citep{ln06,spaans06,wta08,rh09,sbh10,
latif13c,latif13a,schl13}.  In this manner, a 10$^4$ - 10$^5$ \Ms\ Pop III star could be 
formed in less than a Myr.

The evolution of supermassive protostars and stars has been studied with both 
analytical and numerical techniques for over 40 years \citep{fowler1966,wheeler1977,
bond1984,carr1984,fuller1986,bk98,baum99,begel10,hos13,schl13}. Most SMSs are thought 
to become fully convective and then later collapse to massive BHs \citep[e.g.,][]{st79,
shib02,reis13}.  However, previous calculations of SMS evolution have been done at 
low resolution, considered a relatively coarse grid of models in mass, and in some 
cases ignored first-order, post-Newtonian general relativistic corrections to gravity.  
The latter could play an important role in the evolution of 10$^3$ - 10$^5$ \Ms\ stars, 
especially during their pre-explosion phase.  

We have revisited the evolution of Pop III SMSs with high-resolution one-dimensional 
(1D) simulations with updated nuclear reaction rates \citep{heger2001, heger2002} 
and post-Newtonian corrections to gravity.  We have found a 55,500 \Ms\ star that 
explodes as a highly energetic thermonuclear supernova (SN) instead of collapsing to 
a BH \citep[see also][]{montero12}. Our discovery suggests that there may be a range 
of high-mass stars that can die as SNe.  In Section 2 we describe our stellar evolution 
model.  The explosion of the star is examined in 1D in Section 3 and in two dimensions 
(2D) in Section 4.  We discuss the potential impact of this explosion on its host galaxy 
and its visibility at high redshift in Section 5. 

\section{Numerical Model}

\lSect{gsn_Methodology}

The SMS is evolved from the beginning of the main sequence to the onset of collapse 
and then explosion in 1D in \KEPLER{}.  To verify these processes in a separate code, 
and to capture the violent fluid instabilities that can occur during the SN, we also model 
the collapse and explosion of the SMS in 2D in \CASTRO{} with initial conditions taken 
from \KEPLER{}.  Combining 1D and 2D simulations in this manner allows us to 
realistically simulate all phases of the SMS within a practical computational budget.

\subsection{\KEPLER}
\lSect{code}

\KEPLER{} \citep{kepler} is a 1D Lagrangian hydrodynamics and stellar evolution code.  
It includes nuclear burning and mixing due to convection. For a given  stellar mass, 
the initial profiles of density and temperature of a star are determined by solving
 the Lane-–Emden equation \citep{chand1939}. We use the 19-isotope 
APPROX nuclear reaction network \citep{kepler,timmes1999}, which includes heavy-ion
reactions, alpha-chain reactions, hydrogen burning cycles, photo-disintegration of heavy 
nuclei, and energy loss through thermal neutrinos.  Nuclear burning is self-consistently
coupled to hydrodynamics, and we also account for energy deposition due to radioactive 
decay of \Ni{} $\to$ \Co{} $\to$ \Fe{}.   \KEPLER{} uses the Helmholtz equation of state 
\citep[EOS;][]{timmes2000}, which includes contributions from degenerate and 
non-degenerate relativistic and non-relativistic electrons, electron-positron pair production, 
and radiation. The star is partitioned into 1201 zones in mass for an effective resolution of 
46.2 \Ms.  Consistent with the usual convention that massive Pop III stars lose little of their
mass over their lives \citep{Kudritzki00,Vink01,Baraffe01,kk06,Ekstr08}, we turn off mass 
loss in our models.

When stars become extremely massive, $\gtrsim$ 1000 \Ms, general relativity (GR) must 
be taken into account in gravity in stellar evolution models.  Instead of full GR, we apply 
first-order post-Newtonian GR corrections to gravity in our calculations.  We use the 
Tolman-Oppenheimer-Volkoff equation for hydrostatic equilibrium in GR to calculate the 
pressure $P$ \citep{zeldov1971,kippen1990}:
\begin{equation} \label{eqn:tov}
\frac{dP}{dr} = -\frac{Gm}{r^{2}}\varrho (1+\frac{P}{\varrho c^{2}})(1+\frac{4\pi r^3 P}{m c
^{2}}) (1-\frac{2Gm}{r c^{2}})^{-1}, \quad
\end{equation} 
where $r$ is the radius of star, $c$ is the speed of light, $G$ is the gravitational constant, 
and $m$ is the enclosed relativistic mass at $r$, which is the rest mass plus total energy
(internal plus gravitational) divided by $c^2$. The relativistic density $\varrho$ is therefore 
$\varrho_0 +U/c^2$, where $\varrho_0$ and $U$ are the rest-mass and total energy 
densities. When $c^2 \rightarrow  \infty$, Equation (\ref{eqn:tov}) reduces to Newton's 
Law,
\begin{equation}\label{eqn:newton}
\frac{dP}{dr} = -\frac{Gm}{r^{2}}\varrho. \quad
\end{equation}
The effective gravity $\tilde{g}$ is calculated from
\begin{equation}\label{eqn:tov2}                                                      
\tilde{g} = -\frac{Gm}{r^{2}} (1+\frac{P}{\varrho c^{2}})(1+\frac{4\pi r^3 P}{m c^{2}}) (1 - 
\frac{2Gm}{r c^{2}})^{-1}.  \quad 
\end{equation}    
The $U/c^2$ term in $\varrho$ is negligible in our models.  It only becomes comparable to 
the rest energy when $T$ becomes $\sim$ 10$^{13}$ K ($kT \sim m_{\rm p} c^2$, where 
$m_{\rm p}$ is the proton mass).  Equation (\ref{eqn:tov2}) therefore becomes
\begin{equation}\label{eqn:tov3}                                                      
\tilde{g} = g (1+\frac{P}{\varrho_0 c^{2}}+\frac{4\pi r^3 P}{m c^{2}}+\frac{2Gm}{r c^{2}}), 
\quad 
\end{equation} 
where $g$ is the Newtonian gravity field, which is determined from the gravity equation 
in \KEPLER{} and from the monopole approximation to gravity for nearly spherical matter 
distributions in \CASTRO{}.

\subsection{\CASTRO}

\CASTRO{} is a multidimensional adaptive mesh refinement (AMR) hydrodynamics code. 
It has a second-order unsplit Godunov hydro scheme and block-structured AMR \citep{
ann2010,zhang2011}.  We use the same Helmholtz EOS, APPROX	reaction network, 
energy deposition due to \Ni\ and GR corrections to pressure and gravity in \CASTRO{} 
as in \KEPLER{}.  As in \KEPLER{}, this reaction network is sufficient to capture the key 
nucleosynthetic pathways and energetics of SMS evolution and explosion.  \CASTRO{} 
evolves mass fractions for each isotope with its own advection equation.  The monopole 
approximation is used for self-gravity, in which a 1D gravitational potential is constructed 
from the radial average of the density and then applied to gravitational force updates 
everywhere in the AMR hierarchy. This approximation is well-suited to the nearly spherical 
symmetry of the star and is quite efficient. 

We port the star from \KEPLER{} onto a 2D cylindrical coordinate grid in $r$ and $z$ in 
\CASTRO{} with the conservative mapping scheme of \citet{chen2014}.  This is done at 
the onset of collapse, 600 seconds before the core of the star reaches maximum 
compression and the most violent burning begins.  We simulate the half star in 2D, 
so the mesh is 2 $\times$ 10$^{13}$ cm in $r$ and 4 $\times$ 10$^{13}$ cm in $z$, 
which just accommodates the entire star.  
The coarse grid has 256 zones in $r$ and 512 zones in $z$, with up to two levels of 
AMR refinement for an additional factor of up to 16 in resolution. The inner core,
where most of the explosive burning occurs,  is always at the highest resolution;  
additional criteria for mesh refinement depend on gradients of the density, velocity, and pressure.  
The highest spatial resolution of our simulation is about 5 $\times$ 10$^9$cm to resolve the length scale of nuclear 
burning.  Reflecting and outflow boundary conditions are set on the inner and outer boundaries 
in both $r$ and $z$, respectively. 
%We also set up nested grids ensures that the inner core of the 
%star always has the highest resolution, where most of explosive burning occurs. 
The simulation is halted when the SN ejecta become frozen in mass coordinates (homologous expansion). 

\section{The 1D \KEPLER{} Explosion}

\lSect{gsn_results}
The SMS has a radius of $1.73 \times 10^{13}\,\cm$, an effective surface temperature of about 
7.19 $\times$ 10$^{4}$ K, and a luminosity of $5.70 \times 10^{42}\,\erg\,\sec^{-1}$ during 
its pre-collapse phase. The star evolves about 1.69 Myr before beginning to collapse. 
When it begins helium burning, the central density and temperature are $\sim$ 10 $\gcc$ 
and 2.0 $\times$ 10$^{8}$ K, respectively.  Radiation dominates the pressure in the core, 
and the adiabatic index, 
$\gamma_{ad}$, is slightly above 4/3.  At this density and temperature, even the most
energetic photons in the Maxwellian distribution cannot create electron-positron pairs and 
cause $\gamma_{ad}$ to fall below 4/3.  Pair production therefore does not trigger collapse
in the SMS.  Instead, the energy density of the thermal photons in the high temperature
and very low density of the core begins to affect the gravitational field by contributing to the 
source term in Einstein's field equation, in effect becoming an additional source of gravity. 
The ratio of the radiative pressure to the relativistic density $P/(\rho c^2)$ becomes 
$\sim$ 1-3 $\times$ 10$^{-3}$.  This causes $\gamma_{ad}$ in the core to fall below 4/3, 
and it begins to contract. 

\begin{figure}[h]
\begin{center}
\includegraphics[width=\columnwidth]{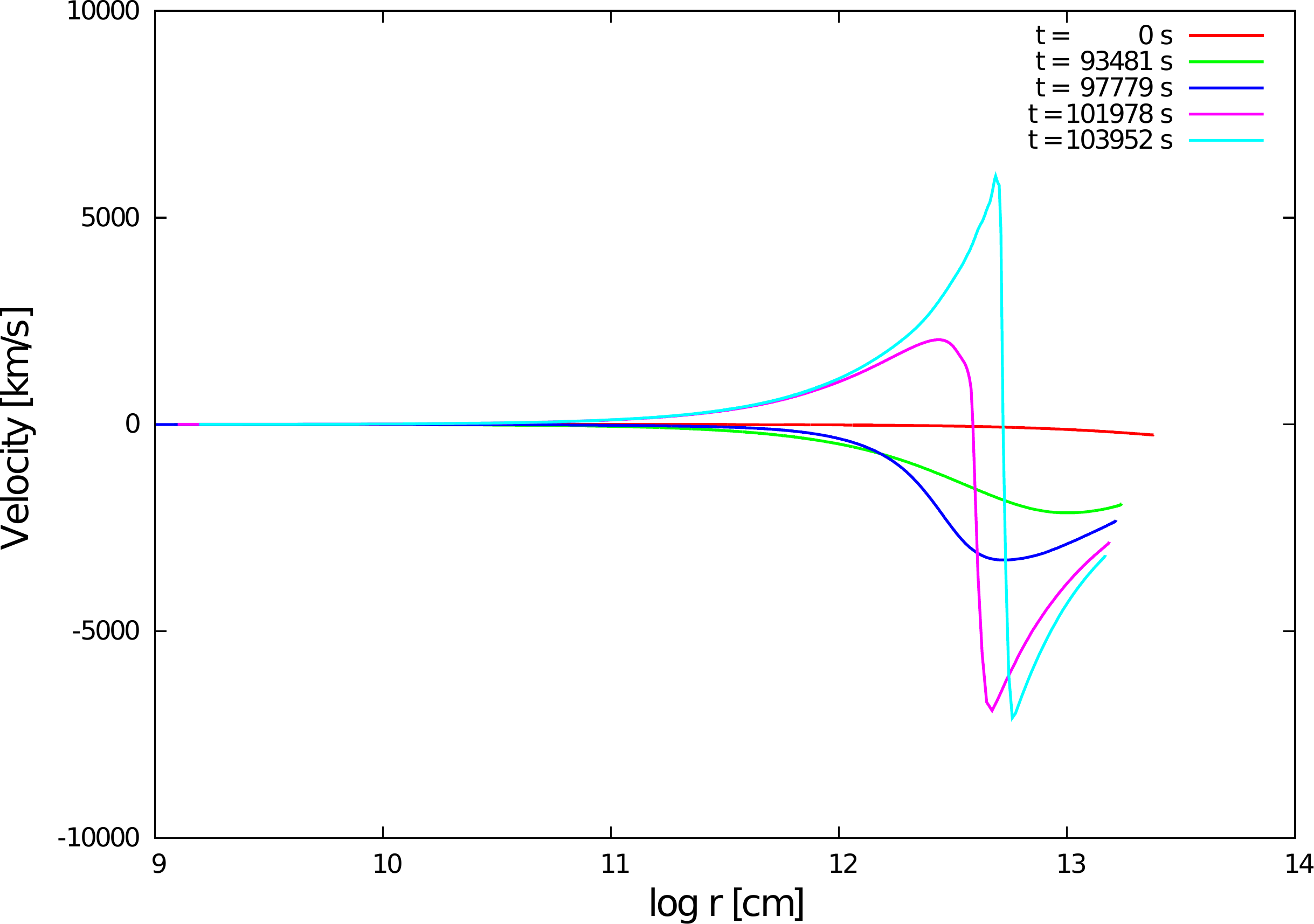} 
\end{center}
\caption[]{Radial velocities in the star from the onset of collapse to explosion. Infall during 
collapse reaches velocities above 7000 km sec$^{-1}$.     
\label{fig:1Dvel}}
\end{figure}

Although the temperature and density in the core now rise rapidly and accelerate nuclear 
burning, they are not high enough to ignite carbon burning, which is the next stage after 
helium burning.  Helium instead begins to burn explosively in the core, burning to carbon 
first through the triple $\alpha$ reaction and then to \Ox{}, \Ne, \Mg and \Si{} through 
$\alpha$ capture reactions.  These reactions release enough energy to reverse collapse. 
We show the evolution of radial velocities in the star in Figure~\ref{fig:1Dvel}.  The onset 
and reversal of collapse is evident, together with the formation of the outgoing shock, which 
propagates outward at several thousand kilometers per second. The shock breaks through 
the surface of the star 11,880 seconds after core bounce.  

\begin{figure}[h]
\begin{center}
\includegraphics[width=\columnwidth]{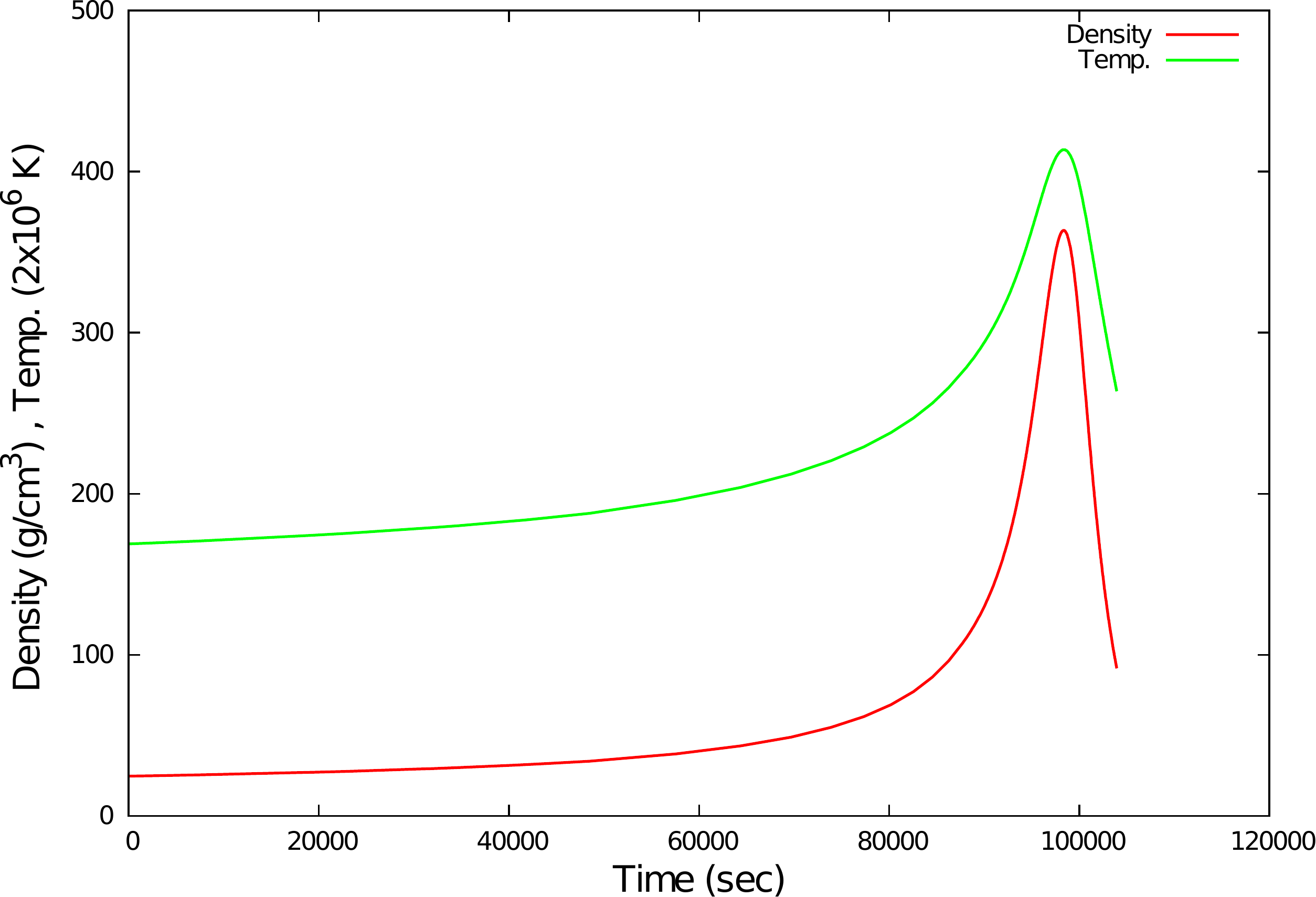} 
\end{center}
\caption{Evolution of central temperature and density during collapse and explosion.  In 
the final 10,000 seconds they reach maximum values of 360 $\gcc$ and 8.26 $\times$ 10$
^8$ K, respectively, before dropping rapidly.  \label{fig:1Dcen}}
\end{figure}

The binding energy of the star prior to collapse is 5.76 $\times$ 10$^{53}$ erg, but 
explosive helium burning releases 6.52 $\times$ 10$^{54}$ erg.  We show the evolution 
of density and temperature at the center of the star in Figure~\ref{fig:1Dcen}.  They are 
initially $\sim$ 20 $\gcc$  and 3.6 $\times$ 10$^{8}$ K, and then rise to peak values of 360 
$\gcc$ and 8.26 $\times$ 10$^8$ K at core bounce before beginning to fall.  The relative 
magnitudes of the binding and explosion energies together with the rapidly falling density
at the center of the star strongly suggest that the SN completely unbinds the star, with no 
BH formation.

\section{The 2D \CASTRO{} Explosion}

\begin{figure}[h]
\begin{center}
\includegraphics[width=\columnwidth]{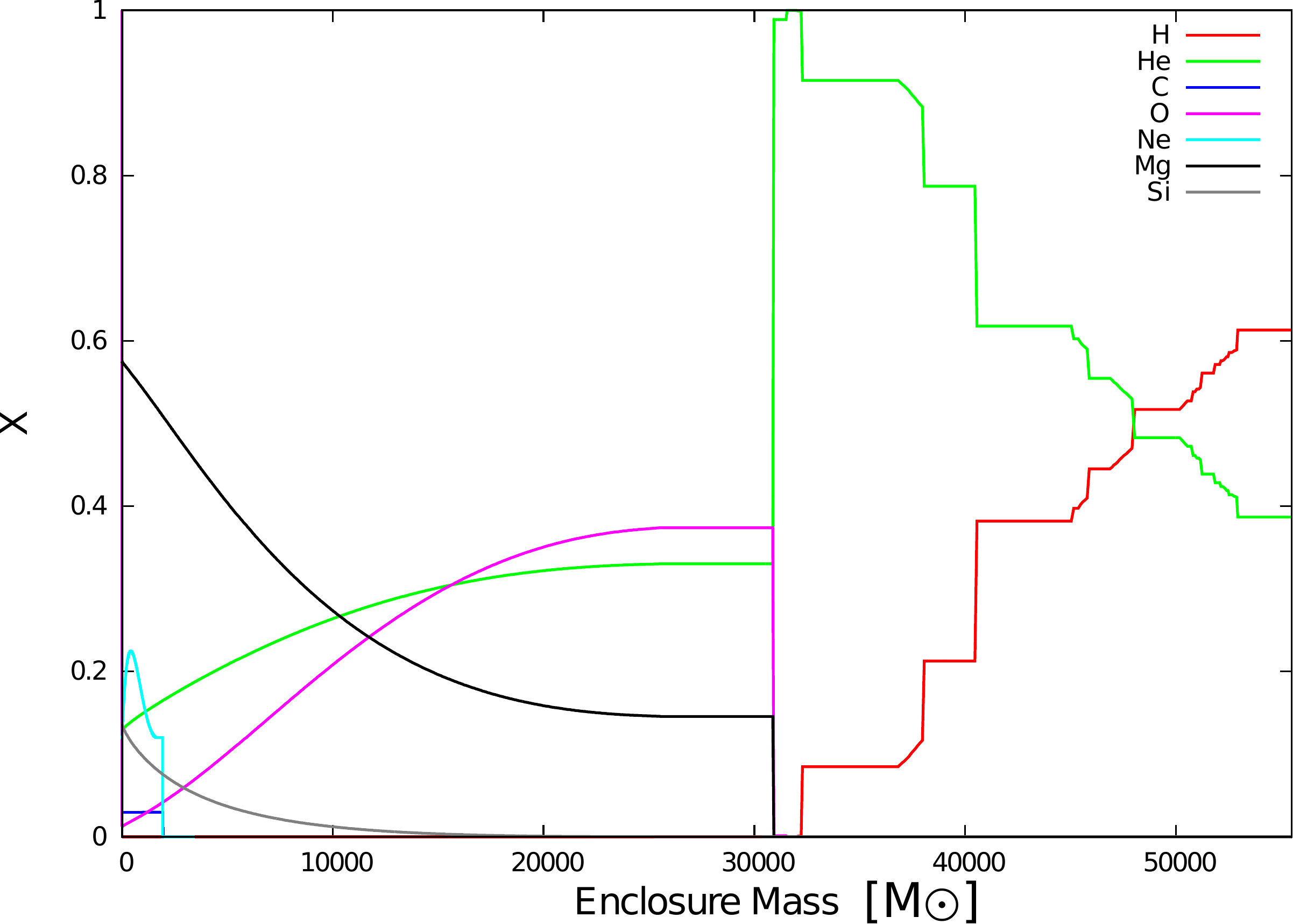} 
\end{center}
\caption[]{Elemental masses at the beginning of the \CASTRO{} simulation. The mass of the 
helium core was about 31,000 \Ms.  Helium abundances at the center of the star are fairly large, 
which implies that central helium burning is still in progress when the star collapses.  
Hydrogen shell burning is also still going on at the base of the envelope of the star.     
\label{fig:mfrac}}
\end{figure}

We initialize \CASTRO{} with the \KEPLER{} profile of the collapsing star 600 seconds 
prior to maximum core compression. In the beginning of the \CASTRO{} run, the core temperature 
$T_c$ and density $\rho_c$ are 8.21 $\times$ 10$^{8}$ K and and 356 $\gcc$, respectively. 
Element masses as a function of radius are shown in Figure~\ref{fig:mfrac}.  The mass of the 
helium core is $\sim$ 31,000\Ms, and it also contains \Si, \Mg, \Ne, \Ox\ and \Cx.  As in 
\KEPLER{}, the collapse is reversed by nuclear energy release, but violent fluid instabilities now 
arise in the core during the explosion.  Helium burning creates pressure gradients that are 
opposite to the gradients of density and mass fraction, which in turn gives rise to Rayleigh-Taylor 
instabilities.  These instabilities are more violent than those in non-reactive flows because 
they mix the fuel with hot ash that then quickly burns the fuel, which in turn enhances the 
pressure gradient and drives more mixing.  This explosion released 8.82 $\times$ 10$^{54}$ erg, 
slightly more than in the 1D model because of the additional burning due to mixing.  

\begin{figure}[h]
\begin{center}
\includegraphics[width=.5\columnwidth]{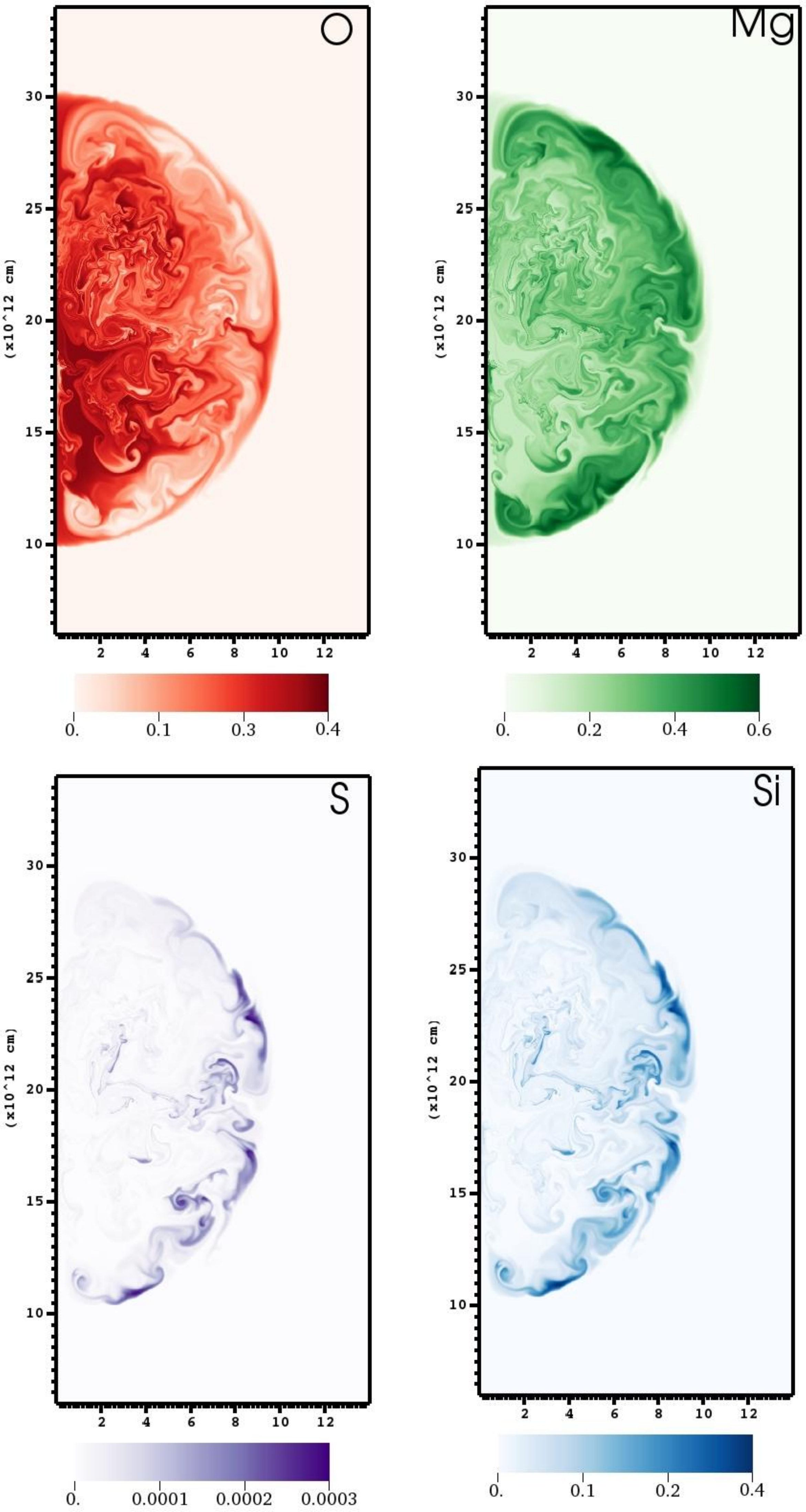} 
\end{center}
\caption{Mixing in \Ox, \Mg, \Si, $\&$ \Sx{} prior to shock breakout.  Mixing on these scales 
might be manifest in the spectra of the explosion.  Few isotopes with atomic numbers larger 
than \Si, such as \Sx, are formed. \label{fig:big_mixing}}
\end{figure}

Burning continues until the shock reaches the hydrogen envelope. Table~\ref{tbl:gsn_table} 
shows elemental abundances before and after the explosion.  Energy release in this SN is
primarily due to the burning of helium, oxygen and neon, which consumes 757 \Ms, 1116 
\Ms, and 291 \Ms, respectively.  Burning in turn yields 1195 \Ms\ of \Mg{} and 970 \Ms\ of 
\Si.  Few isotopes beyond \Si{} are synthesized, with $\ll$ 1 \Ms\ of \Ni{} being formed, not 
enough to be detected.  Since the hydrogen envelope of the SMS is not as extended as in 
red supergiants, the shock does not plow up much mass when it crashes into it. The reverse 
shock is therefore weak and does not drive further mixing.  We show the degree to which \Ox, 
\Mg, \Si, $\&$ \Sx{} have mixed by the time the shock has broken out of the surface in Figure
\ref{fig:big_mixing}.  Most mixing has now ceased.  The major drivers of mixing in this model
are fluid instabilities that emerge during burning.  Mixing can cause heavy elements to be
dredged up from deeper layers and appear in the SN spectra at early times. 

In \KEPLER{} models that exclude GR, the $55,500\,\Msun$ star collapses to a BH instead of
exploding, {\color{black} because its core continues evolving through helium burning then 
encounters the pair instability before it ignites carbon burning. Once the collapse of core 
triggered by pair instabilities happens, the nuclear burning is not able to halt  the 
collapse at this time and the entire star eventually dies as a black hole.}  
Likewise, if the star is 56,000 \Ms\ it collapses to a BH, even when GR is included.   
This suggests that there is a narrow window in mass around 55,500 \Ms\ for the progenitor 
masses of general relativistic supernovae (GSNe).  More masses need to be tested, and more 
extensive nuclear reaction networks may be needed to capture all the nucleosynthetic pathways 
in GSNe.  Our \KEPLER{} and \CASTRO{} simulations show that GSNe are only weakly 
dependent on dimensionality.  

GSNe are different from pair-instability SNe 
\citep[PSNe;][]{heger2002,scannapieco2005,chen2014b} in every aspect:  
what triggers the collapse, 
what drives the explosion, the degree of mixing, and \Ni{} production.  We compare properties 
of PSNe and GSNe in Table~\ref{tbl:gsn_table2}.  In GSNe, only a trace amount of \Ni{} ($\ll$ 1 
\Ms) is produced at the edge of the oxygen-burning shell by $\alpha$ capture.  PSNe can 
synthesize up to 50 \Ms\ of \Ni{} through explosive \Si{} burning.  The light curves of GSNe are 
mainly powered by the thermal emission by hot ejecta and the conversion of kinetic energy into 
entropy by the shock, not \Ni{} decay.  Large masses of elements with atomic masses between 
\Cx{} and \Si{} are synthesized during the explosion, and they are all dispersed into the 
surrounding medium.

\section{Discussion and Conclusion}

\lSect{gsn_conclusions}

We have discovered that Pop III stars with masses of 55,500 \Ms\ may explode as SNe 
instead of collapsing to BHs. 
GR effects, rather than the pair instability, trigger the explosion, which at $\sim$ 9 $\times$ 
10$^{54}$ erg is the most energetic thermonuclear SN known. Mixing in 2D enhances burning 
and thermonuclear yields during the explosion, which completely unbinds the star and leaves 
no compact remnant.  Energy release is primarily due to explosive helium burning after the 
onset of central helium burning. The explosion yields mostly elements between carbon and 
silicon, with almost no iron group elements.  Besides enhancing burning, mixing during the SN 
can dredge up heavy elements from deeper layers and cause them to appear at earlier times 
in spectra. {\color{black}  \citet{montero2012} recently studied the relativistic 
collapse of SMS and discovered that explosions of SMS occurred even 
for stellar masses up to $5\times10^{5}$ \Ms, which are powered by the
 hydrogen burning due to hot CNO cycle. 
In their simulations, they use simplified stellar models, with a single fluid of fixed hydrogen, 
helium, and CNO metal abundance. More importantly, all of their exploded models are associated 
non zero metallicity, however, there is no metal inside the Pop~III stars when they are born. 
Hence the hot CNO cycle may not occur for the Pop~III SMS at early on.  Instead, our SMS model 
evolves from the main sequence through collapse to explosion with the detailed stellar physics 
which self-consistently produces CNO from the primordial elements. 
This provides a more realistic pre-supernova model. The differences in pre-supernova models 
as well as other microphysics used can significantly alter the explosion mechanics that may 
explain the discrepancies between their findings and ours. }

Our simulations are approximate for several reasons. First, we do not model the evolution of 
the protostar from much lower masses.  Instead, the star is initialized at the beginning of the 
main sequence in \KEPLER{}.  Next, we did not evolve the star under the heavy ongoing 
accretion that gave birth to the star. We also did not consider stellar rotation, which could lower 
the mass at which the star explodes \citep[e.g.,][]{cw12,cwc13}. There is growing evidence that 
many Pop III stars are born with high rates of spin \citep{stacy11b,get12}.  Finally, we did not 
include UV feedback from the star, which could regulate its accretion rates.  However, in some 
cases luminosity from the star is thought to terminate accretion \citep{johnson2012}, so our 
assumption that the star has a constant mass is plausible.  Efforts are now underway to survey 
Pop III protostellar evolution under a variety of accretion rates and SMS evolution with much 
larger and finer grids in mass.

What effect would such an explosion have on the protogalaxy that gave birth to the star?  The 
answer depends on the ambient density of the SN.  If the SMS grows by accretion through a
disk its UV radiation will break free of the disk and ionize the surrounding envelope.  In these
circumstances the star would explode in low densities, and \citet{jet13a} show that the SN 
would drive all the gas from the protogalaxy, perhaps engulfing nearby protogalaxies with
metals.  Much of this material would later fall back to the halo on timescales of 50 - 100 Myr
and form stars.  If accretion is spherical, then the star will not ionize its envelope and it will
explode in very high densities \citep{johnson2012,hos13,schl13}.  In this scenario, \citet{wet13b,wet13a} find 
that much of the energy of the explosion is promptly radiated away by bremsstrahlung X-rays 
and inverse Compton scattering of cosmic microwave background (CMB) photons, but that the 
SN remnant still expands to roughly the virial radius of the halo, $\sim$ 1 kpc, and then 
collapses back into the halo.  Collapse thoroughly mixes baryons in the halo with metals and 
may trigger a starburst that would easily distinguish this protogalaxy from its dimmer and less 
rapidly evolving neighbors.  

Radiation hydrodynamical simulations show that the GSN would be visible at $z \lesssim$ 
20 to future all-sky NIR surveys by {\it Euclid}, the Wide-Field Infrared Survey Telescope 
(WFIRST) and the Wide-field Imaging Surveyor for High-Redshift (WISH) \citep{wet12d}.  
The wide survey fields and high sensitivities of these missions would enable them to detect 
such an event anywhere in the universe, even if their numbers are small.  Furthermore, a 
single GSN can create 100 times the chemical yield of a PSN.  But unlike PSNe, which 
synthesize more iron group elements, GSNe would mainly enrich primordial gas with 
elements from \Cx{} to \Si.  Traces of GSNe might  therefore be found in early galaxies 
that are \Fe{} deficient but enhanced with \Cx{} and \Ox{} \citep{keller2014}. Whether in future NIR campaigns 
or in the fossil abundance record, the detection of GSNe in the early universe may soon 
signal the birth of SMBH seeds and the first quasars.
 
\acknowledgments
 {\color{black} The authors thank the anonymous referee  for reviewing this manuscript and providing 
 insightful comments}, and the members of CCSE at LBNL for help with \CASTRO{}. We also thank 
Volker Bromm, Dan Kasen, Lars Bildsten, John Bell, and Adam Burrows for many useful 
discussions.  K.C. was supported by an IAU-Gruber Fellowship, a Stanwood Johnston 
Fellowship, and a KITP Graduate Fellowship.  A.H. was supported by a Future Fellowship 
from the Australian Research Council (ARC FT 120100363).  D.J.W. was supported by the 
Baden-W\"{u}rttemberg-Stiftung by contract research under the programme Internationale 
Spitzenforschung II (grant P-LS-SPII/18).  All numerical simulations were performed at the 
University of Minnesota Supercomputing Institute and the National Energy Research 
Scientific Computing Center.  This work was supported by the DOE grants DE-SC0010676, 
DE-AC02-05CH11231, DE-GF02-87ER40328, DE-FC02-09ER41618 and by NSF grants 
AST-1109394 and PHY02-16783.  Work at LANL was done under the auspices of the 
National Nuclear Security Administration of the US Department of Energy at Los Alamos 
National Laboratory under Contract No. DE-AC52-06NA25396.

\newpage

\begin{table}[h]
\begin{center}
\begin{tabular}{lcccccccc}
\hline 
\\
Isotope & \Hy &  \He     &      \Cx      &      \Ox     & 	\Ne     &  	\Mg    &   \Si  &  Total \\
\\
\hline
{ } &  $\Msun$  &   $\Msun$  & $\Msun$  &   $\Msun$ & $\Msun$  &   $\Msun$ & $\Msun$  &   $\Msun$ \\
Before &  8336  &  24902 &  922 & 7972 & 5110 & 7748 & 515 &  55505             \\
After  &   8335 &  24145 & 919 & 6856 & 4819 & 8943 & 1485 & 55502                   \\
$\Delta M$ & -1 &  -757 & -3 & -1116 & -291 & 1195 & 970 &  -3     \\
\hline
\end{tabular}
\caption{Elemental masses before and after the explosion. \label{tbl:gsn_table}}
\tablecomments{ The loss of $3\,\Msun$ in total is due to the diffusion of 
the stellar envelope out of the simulation domain, which is only a tiny fraction of 
overall mass. It does not affect the explosion and nucleosynthesis.  }
\end{center}
\end{table}

%\appendix
\begin{table}[h]
\begin{center}
\begin{tabular}{lcc}
\hline 
\\
Characteristic Property  &   PSNe &  GSNe    \\
\hline  \\
Progenitor Mass ($\Msun$)  & $150 - 260$     &  $\sim 55,500$                  \\
Collapse Trigger & Pair Instability     &     GR Instability     \\
Explosive Burning &    \Ox, \Si {}   &\He{}               \\
\Ni{} Production  ($\Msun$)  &   $0.1 - 50$  & $\ll 1$\\
Explosion Energy ($\erg$) & $1-100 \times 10^{51}$     &   $6-9 \times10^{54}$ \\
Fluid Instabilities    &    Reverse Shock   &   Burning \\
\hline
\end{tabular}
\caption{Comparison between PSNe and GSNe \label{tbl:gsn_table2}}
\end{center}
\end{table}

\end{document}